\documentclass[fleqn,twoside]{article}

\usepackage{espcrc2}

\usepackage{graphicx}

\newcommand{\AmS}{{\protect\the\textfont2
  A\kern-.1667em\lower.5ex\hbox{M}\kern-.125emS}}
\title{Selection of Tau Leptons with the CDF II Trigger System}

\author{S.Baroiant\address[DAV]{Department of Physics,University of California, Davis, CA 95616, USA},
    M.Chertok\addressmark[DAV],
    M.Goncharov \address[TXS]{Department of Physics, Texas A\&M University, College Station, TX 77843, USA},
    T.Kamon\addressmark[TXS],
    V.Khotilovich\addressmark[TXS],
    R.Lander\addressmark[DAV],
    P.Murat\address[Frm]{Fermi National Accelerator Laboratory, Batavia, IL 60510, USA},
    T.Ogawa\address[WSD]{Department of Physics, Waseda University, Tokyo 169, Japan},
    C.Pagliarone\address[PIS]{Istituto Nazionale di Fisica Nucleare,  I56100 Pisa, Italy},
    G.M.Piacentino\addressmark[PIS],
    F.Ratnikov\address[RTG]{Department of Physics and Astronomy, Rutgers, the State University of New Jersey, Piscataway, NJ 08854,
    USA}
    A.Safonov\addressmark[DAV],
    A.Savoy-Navarro\addressmark[Frm]\address[PRS]{LPNHE Universites de Paris 6 et 7/IN2P3-CNRS, Tour 33, RdC 4, Place Jussieu, 75252 Paris Cedex 05, France},
    J.R.Smith\addressmark[DAV],
    D.Toback\addressmark[TXS],
    S.Tourneur\addressmark[Frm]\addressmark[PRS],
    E.Vataga\addressmark[PIS]\thanks{Corresponding author. 
	Tel.: +39-050-2214415; fax: +39-050-2214317; e-mail: Elena.Vataga@pi.infn.it}}

\begin{document}
\begin{abstract}
In Run II of the CDF experiment, traditional dilepton triggers are
enriched by lepton ($e$ or $\mu$) plus track, di-$\tau$ and $\tau$
plus missing transverse energy triggers at Level-3 dedicated to
physical processes including tau leptons.  
We describe
these triggers, along
with their physics motivations, implementation and cross-sections and
report on their initial performance.

\vspace{1pc}

\end{abstract}


\maketitle
\section{Introduction}

Run II of the Tevatron produces $p\bar{p}$ interactions
with an instantaneous luminosity about $10^{32}$ ${\rm
cm^{-2}\,s^{-1}}$.  The CDF Detector trigger system \cite{TDR}
decreases the resulting initial event rate of 2.5 MHz
down to 70 Hz for data storage and offline analysis, with the goal
of efficient extraction of interesting physics events from the
large minimum bias background.  One important subset of these physics
signatures includes $\tau$ leptons. Events with $\tau$'s are
important for the study of various Standard Model (SM) processes
such as $Z\rightarrow \tau\tau$, $t\bar{t}$ with 
$W\rightarrow \tau \nu_{\tau}$, $H\rightarrow\tau\tau$ 
and also to extend the reach of searches for physics
beyond the SM.  These
physics goals drive us to look for $\tau$-like objects starting with
Level-3 of the trigger, and our implementation selects events with
two leptons in the final state: $e\tau_h, \mu \tau_h,
\tau_h\tau_h$ and also $\tau_h\nu$, 
as well as $ee, e\mu$ and $\mu\mu$. With "$\tau_h$" we
indicate hadronically decaying $\tau$'s, while the $e$ or $\mu$ can
be either produced directly or through a leptonic decay of the
$\tau$.

These tau triggers
benefit from the upgraded CDF trigger system, in
particular the tracking processor at Level-1 (L1)  and the refined tracking
information available at Level-2 (L2) \cite{TDR,XFT}.


\section{Implementation of the $\tau$ Triggers}

\subsection{$\tau$ reconstruction}
Taus promptly decay in leptonic or hadronic (65\%) modes with at
least one $\nu$ in the final state. Hadronic $\tau$ decays have
the distinct signature of a narrow isolated jet with low
multiplicity (1 or 3 prongs) and low visible mass ($<M_{\tau}$).
The basis for our triggers, therefore, is a $\tau$-cone algorithm
for the reconstruction of these decays. To build the
$\tau$-cone object, we start with a narrow calorimeter cluster,
above a suitable $E_T$ threshold, 
matched to a seed track with momentum above a $P_T$ cut. The region
within an angle $\Theta_{Sig}$ from the seed track direction is
used to define a cone of tracks to be associated with the $\tau$.
The region between $\Theta_{Sig}$ and $\Theta_{Iso}$ defines the
isolation cone: we require that no tracks with $P_T$ higher
than a fixed low threshold be found in this region. At the Level-3 (L3)
of the trigger, $\Theta_{Sig}$ and $\Theta_{Iso}$ have values of
$10^{\circ}$, and $30^{\circ}$, respectively, and are determined in 3
dimensions. 

We have implemented 4 different triggers for physics processes
with $\tau$'s in the final state, installed in the CDF
trigger tables in January 2002.  Below we describe their main
features.

\subsection{Electron plus track, Muon plus track}

The selection for the $e$ plus track trigger starts with L1 and L2
requirements of a single EM tower with $E_T>$ 8 GeV with an associated
XFT track with $P_T>$ 8 GeV/$c$ and a second track with $P_T>$ 5 GeV/$c$.
At L3, these conditions are refined and 
charged track isolation around the reconstructed (2nd) track is
imposed. 
The current average cross-section for this L3 trigger is $\sim$29 nb.
For the muon plus track trigger we require a muon stub matched at
L1 to an XFT track with $P_T>\; 4 GeV/c$, with an increased track
threshold of 8 GeV/$c$ at L2.  The "track" requirements are identical to
those for the electron plus track trigger.  The current average
cross section for this trigger is $\sim$16 nb.

\subsection{Di-$\tau$,$\tau$ plus Missing Transverse Energy}

At L1 the di-$\tau$ trigger requires 2 calorimeter towers with $E_T>$ 5 GeV
and 2 matching XFT tracks with $P_T>$ 6 GeV/$c$, separated by an angle of
$\phi>30^{\circ}$. L2 requires cluster
$E_T>$ 10 GeV and imposes track isolation. At L3, using the
full reconstruction code, 2 $\tau$ candidates with seed track
$P_T>$ 6 GeV/$c$,  and originating from the same vertex: $|\Delta
Z|<10$ cm, are required. The L3 cross section for this trigger is
$\sim$12 nb.

The $\tau$ plus missing transverse energy (MET) trigger
requires $E_{MET}>$10 GeV
at L1. At L2 this is increased to $E_{MET}>$20 GeV and a
calorimeter cluster with an isolated track with $P_T>$10 GeV/$c$
is required. This is followed by full event reconstruction at L3,
requiring at least one $\tau$ candidate having seed track $P_T>$
4.5 GeV/$c$. The present trigger cross section is 5 nb.

\section{First Results}
\begin{figure}[htb]

\includegraphics*[width=18pc]{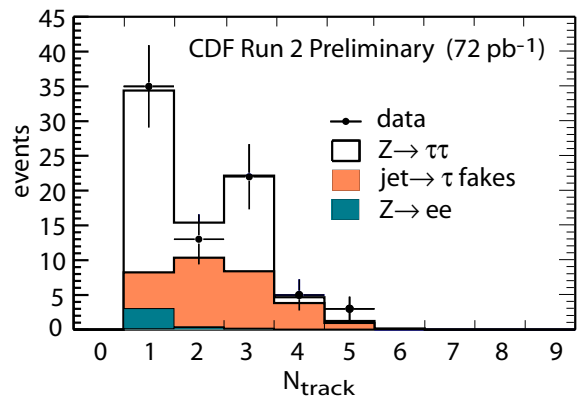}
\caption{Track multiplicity for the $\tau$ candidate.}
\label{fig:ztt}
\end{figure}

The study of $Z\rightarrow \tau_e \tau_h$ is performed using
$72 \;{\rm pb}^{-1}$ of data taken with the electron plus track
trigger \cite{safonov}. The track multiplicity associated with the $\tau_h$
candidate found in events passing selection cuts is shown in
Fig.~\ref{fig:ztt}. We observe a clear $\tau$ signal over
background levels even before requiring an opposite sign charge
for the electron and the $\tau_h$. The mass distribution of the
electron-$\tau_h$-MET system in the data is also found to be
consistent with the $Z\rightarrow \tau\tau$ hypothesis.

%
%
%

\end{document}